\documentclass[english,prd,superscriptaddress,nofootinbib,
               preprintnumbers,twocolumn,showpacs]{revtex4}

\usepackage[latin1]{inputenc}
\usepackage{amsmath}
\usepackage{amssymb}
\usepackage{graphicx}
\usepackage[caption=false]{subfig}
\usepackage{amsthm}
\usepackage{bm}
\usepackage{color}
\usepackage{amsfonts}
\usepackage{dcolumn}
\usepackage{babel}
\usepackage{amssymb}
\usepackage{epsfig}
\usepackage{graphics,psfrag,rotating}
\usepackage{dcolumn}
\usepackage{bm}
\usepackage{epstopdf}
\usepackage{color}
\usepackage[usenames,dvipsnames,svgnames]{xcolor}
\usepackage[colorlinks=true,
            linkcolor=red,
            urlcolor=gray,
            citecolor=blue]{hyperref}
  \usepackage{hyperref}

\makeatletter

\def\be{\begin{equation}}
\def\ee{\end{equation}}
\def\ba{\begin{eqnarray}}
\def\ea{\end{eqnarray}}

\newcommand{\arXiv}[2][]{\href{http://arxiv.org/abs/#2}{\texttt{arXiv:#2\@ifempty{#1}{}{ [#1]}}}}

\begin{document}
\title{On the Emergence of Accelerating Cosmic Expansion \\in f(R) Theories of Gravity}
\author{Timothy Clifton}
\affiliation{School of Physics and Astronomy,
 Queen Mary University of London, 
Mile End Road, London E1 4NS, UK.}

\author{Peter K. S. Dunsby}
\affiliation{Astrophysics Cosmology \& Gravity Center, and Department of Mathematics \& Applied Mathematics,
University of Cape Town, 7701 Rondebosch, South Africa.}
\affiliation{South African Astronomical Observatory,  Observatory 7925, Cape Town, South Africa.}

\begin{abstract}
We consider cosmological modelling in $f(R)$ theories of gravity, using both top-down and bottom-up constructions. The top-down models are based on Robertson-Walker geometries, and the bottom-up constructions are built by patching together sub-horizon-sized regions of perturbed Minkowski space. Our results suggest that these theories do not provide a theoretically attractive alternative to the standard general relativistic cosmology. We find that the only $f(R)$ theories that can admit an observationally viable weak-field limit have large-scale expansions that are observationally indistinguishable from the Friedmann solutions of General Relativity with $\Lambda$. Such theories do not alleviate any of the difficulties associated with $\Lambda$, and cannot produce any new behaviour in the cosmological expansion without simultaneously destroying the Newtonian approximation to gravity on small scales.
\end{abstract}
\maketitle
\section{Introduction}

As we approach its centenary \cite{Einstein},  General Relativity (GR) remains our most promising theory for describing the gravitational interaction.  It has had tremendous success in describing relativistic gravity in the solar system, and in binary pulsar systems \cite{will}. When applied to cosmology, it has resulted in the $\Lambda$CDM (or concordance) model of the Universe \cite{concordance}. This model has the benefit of extreme simplicity, being based on a single universal Robertson-Walker geometry. It has also been extremely successful at reproducing a wide variety of cosmological observations, including supernovae \cite{sneIa}, CMB anisotropies \cite{cmbr}, large-scale structure formation \cite{lss}, baryon oscillations \cite{bo}, and weak lensing \cite{wl}. However, the $\Lambda$CDM model appears to necessitate the largest fine-tuning problem that has ever occurred in physics, and the apparent coincidence that we live at the exact time when the cosmological constant is just starting to dominate the large-scale expansion \cite{cc}. These problems have led many authors to consider alternative theories of gravity \cite{reviews}, often based on modifications to the usual Einstein-Hilbert action that are expected to become large at low curvatures.

One of the simplest ways to extend GR is to consider gravitational theories that have field equations that are derived from the metric variation of a Lagrangian density that is a non-linear function of $R$: The so-called $f(R)$ theories of gravity. Such theories have the pleasing property of being constructed entirely from the metric tensor, are well-motivated from attempts to improve the renormalizability of gravity \cite{stelle}, and have often been considered in early universe physics \cite{star80}. They have also been particularly well studied in recent years, as they potentially allow the late-time accelerating expansion of the universe to be interpreted as having a geometric origin, rather than being due to a violation of the energy conditions in the matter sector.

Despite this wealth of work, very little effort has so far gone into understanding how weak-field systems can be embedded into cosmological solutions within this class of theories. Exact inhomogeneous cosmological solutions are very difficult to find, and Swiss-cheese-type constructions have only recently been considered \cite{ref1}. In principle, however, the cosmological behaviour should in some way determine the boundary conditions for the weak-field systems on small scales, and the cosmological expansion should be able to be viewed as being due to bodies on small scales falling away from each other. These ideas have been made explicit in General Relativity, where there exists a deep relationship between Newtonian force laws proportional to $r^{-2}$ and $r$, and the dust-dominated and $\Lambda$-dominated Friedmann solutions \cite{Clifton}. It is the purpose of the current paper to consider the corresponding situation in $f(R)$ theories of gravity that lead to accelerating expansion at late times.

We begin by considering the Friedmann solutions of these theories, and demonstrate that there is great freedom in the large-scale evolution of homogeneous and isotropic spaces, as long as one is allowed complete freedom in choosing $f(R)$. We then go on to consider how some particular theories, that have an evolution very close to $\Lambda$CDM today, generically exhibit growing oscillations that lead to curvature singularities at finite times. In contrast, we then consider the cosmological models that emerge from joining small regions of perturbed Minkowski space together using the appropriate junction conditions. The geometry inside each of the regions is expanded as a power series in the weak-field slow-motion limit. We find that the only theories that admit a viable weak-field limit within each of these regions are also forced to expand like $\Lambda$CDM on large scales.

This result has several interesting implications. Firstly, it suggests that the freedom-in-principle that exists in the Friedmann solutions of $f(R)$ theories of gravity does not exist if one limits oneself to observationally viable theories. Secondly, it suggests that the potentially problematic oscillations that can occur in their Friedmann solutions do not exist if the space-time can be described as slowly-varying weak-field perturbations about Minkowski space on small scales. Thirdly, it shows that viable $f(R)$ theories of gravity cannot solve any of the problems associated with the cosmological constant, or generalize its evolution in any way: Their large-scale evolution is identical to $\Lambda$CDM, and the effective value of $\Lambda$ must be constructed from parameters that appear in the gravitational action only. In this sense, there is only a nominal difference between late-time accelerating expansion in GR with $\Lambda$, and in viable $f(R)$ theories.

In this paper, we start in Section \ref{sec:pre} by introducing the $f(R)$ theories of gravity, as well as the mathematical tools that will be required build our models. In Section \ref{sec:flrw} we then continue to explore the freedom that exists in the FLRW models that are constructed within these theories, as well as the instabilities that can exist within an important class of them that is thought to be observationally viable. In Section \ref{sec:pn} we then go on to consider cosmological models built from the bottom up. we find that the models that are built in this way, and that have an observationally viable weak-field limit, have severely constrained large-scale expansions. We summarize and conclude in Section \ref{disc}.

\section{Preliminaries}
\label{sec:pre}

The $f(R)$ theories of gravity are a family of action-based theories, whose Lagrangian density is given by
\begin{equation}
\mathcal{L}= \frac{1}{16 \pi} f(R)+{\cal L}_{m} \;,
\label{action}
\end{equation}
where $R$ is the Ricci scalar, $f(R)$ is a general differentiable (at least $C^2$) function of the Ricci scalar, 
and $\mathcal{L}_m$ corresponds to the Lagrangian of standard matter fields. Here, and throughout, we use units in which $c=G=1$.

Variation of the action $\mathcal{A} = \int d^4x \sqrt{-g} \mathcal{L}$ with respect to the metric, $g_{\mu \nu}$, yields the following gravitational field equations:
\be
\label{field}
f'R _{\mu\nu} - \frac12 f g_{\mu\nu} - \nabla_{\nu}\nabla_{\mu}f' + g_{\mu\nu} \nabla_{\sigma} \nabla^{\sigma} f' = 8\pi T_{\mu\nu},
\ee
where we used $f=f(R)$, and where primes denote derivatives with respect to $R$. The energy-momentum tensor derived from varying the Lagrangian associated with standard matter fields is denoted by $T_{\mu \nu}$.

The Lagrangian (\ref{action}) and field equations (\ref{field}) are known to contain an additional scalar degree of freedom, that is not present in GR. This can be made explicit by conformal transformation of the metric \cite{Whitt:1984pd, Maeda:1988ab}, or alternatively by a Legendre transformation of the theory. We will return to this equivalency in Section \ref{disc}, but for the majority of the paper we will choose to keep the theory written as a function of the metric only, as in Eqs. (\ref{action}) and (\ref{field}).

Let us now consider the tools required for our top-down and bottom-up constructions: That is, the Friedmann equations and the post-Newtonian expansion.

\subsection{Friedmann Equations}

The Friedmann equations are derived using the homogeneous and isotropic Robertson-Walker geometry:
\be
\label{ds}
ds^2 = -dt^2 + a^2(t) \left[ \frac{dr^2}{1-k r^2} +r^2 (d\theta^2 +\sin^2 \theta d \phi^2) \right] \, ,
\ee
where $a(t)$ is the scale factor, and where $k$ is the (constant) curvature of space.

Substituting Eq. (\ref{ds}) into the field equations (\ref{field}), and taking the matter source to be a perfect fluid, gives the {\em Friedmann} (constraint) equation as
\begin{equation}\label{fried}
3H^2=8\pi\rho+\rho_R \, ,
\end{equation}
where
\begin{equation}
\rho_R=3H^2(1-f')+\frac{Rf'-f}{2}-3H f'' \dot{R}
\end{equation}
is the conserved energy density of the effective $f(R)$ dark energy, $H \equiv \dot{a}/a$ is the Hubble parameter, and $\rho$ and $p$ are respectively the energy density and isotropic pressure of the perfect fluid (corresponding to standard matter fields). The {\em Raychaudhuri} (evolution) equation is
\ba
\label{ray}
3\dot{H}+3H^2 &=&-\frac{1}{2f'} \Big[ 8 \pi (\rho+3p)+f-f'R   \\&\;& \qquad \qquad +3H f''\dot{R}+3f'''\dot{R}^2+3f''\ddot{R}\Big]\;. \nonumber
\ea
The energy density and pressure of the matter fields are related by the energy conservation equation:
\begin{equation}\label{cons:perfect}
\dot{\rho}+3 H\left(\rho+p\right) =0\;,
\end{equation}
while $\rho_R$ satisfies
\begin{equation}
\dot{\rho}_R+3H(\rho_R+p_R)=0\;,
\end{equation}
where 
\begin{equation}
p_{R}=\ddot{f'}+2H\dot{f'}+\frac{f-f'R}{2}+\left( H^{2}-\frac{R}{3} \right)(1-f')\;.
\end{equation}
The effective equation of state for $f(R)$ dark energy is then given by 
\begin{equation}
w_R \equiv \frac{p_R}{\rho_R}\;.
\label{DE}
\end{equation}
Here, and throughout what follows, we limit ourselves to the spatially flat geometries in which $k=0$. The other cases follow straightforwardly.

The solutions to Eqs. (\ref{fried})-(\ref{cons:perfect}) are the basis of the Friedmann-Lema\^{i}tre-Robertson-Walker (FLRW) cosmological models, which will be discussed in Section \ref{sec:flrw}.

\subsection{Post-Newtonian Expansion}
\label{pn}

The post-Newtonian expansion is a weak-field, slow-motion expansion about a Minkowski space background. It is the basis of almost all modern investigations of relativistic theory in weakly-gravitating systems \cite{will}, and has been shown to have an almost unreasonably wide field of applicability \cite{will2}. The expansion is based on assigning all quantities (geometric, kinetic and matter) orders of smallness in the parameter $\epsilon=v/c$, where $v$ is the typical relative velocity of the matter fields that fill the space-time, and $c$ is the speed of light. For the solar system, galaxies and galaxy clusters this parameter is expected to span the range $10^{-5}$--$10^{-3}$.

The geometry for the lowest order part of the post-Newtonian expansion can be written as
\be
\label{pnmetric}
ds^2 = -(1-2 \phi) dt^2 +(1+2 \psi) dx_i dx^i +2 A_i dt dx^i,
\ee
where $\phi$, $\psi$ and $A_i$ are all gravitational potentials.  The lowest order parts of the energy-momentum are then taken to be
\ba
T_{00} &=& \rho \label{t1} \\
T_{0i} &=& -\rho v_i \label{t2}\\
T_{ij} &=& p \delta_{ij} \, , \label{t3}
\ea
where $i$ and $j$ run over spatial indices, and $\delta_{ij}$ is the 3-dimensional Kronecker delta.

For ordinary matter content, the orders-of-smallness of the quantities in Eqs. (\ref{pnmetric})-(\ref{t3}) are given by
\be
\phi \sim \psi \sim \epsilon^2 \, ,
\ee
and
\ba
v_i &\sim& \epsilon \\
A_i &\sim& \epsilon^3 \\
\frac{p}{\rho} &\sim& \epsilon^2 \, .
\ea
Each and every time derivative is also taken to add an extra order-of-smallness to the quantity that it operates on, with respect to spatial derivatives, so that
\be
 \frac{{\partial }/{ \partial t}}{{\partial }/{ \partial x}} \sim \epsilon \, .
\ee
These quantities are all dimensionless, making an assignment of an order of magnitude to them unambiguous. In the field equations, however, the terms that appear have dimensions of one-over-length-squared. The best one can do for such terms is therefore to say that they have an order-of-smallness when expressed in some choice of units for length. We choose these units such that spatial derivatives add no order-of-smallness to the objects on which they operate, which results in $\rho \sim \epsilon^2$. This choice is for book-keeping purposes only, and does not affect any of the resulting equations. Any other choice of units would result in an identical set of equations.

Of course, any additional fields that appear in the modified theories of gravity, such an extra scalar in the case of $f(R)$ theories, must also be expanded in powers of $\epsilon$, if the theory is to fit into the post-Newtonian framework. The procedure for solving Einstein's equations (or the field equations of the theory in question, if different to GR) is then to solve each equation to lowest non-trivial order in $\epsilon$, before moving on to higher orders. 

The lowest orders in the diagonal components of Einstein's equations are given to even orders in $\epsilon$, while the lowest order in the off-diagonal components are odd in $\epsilon$. The Newtonian limit of the equations of motion for massive particles is then given in terms of the $O(\epsilon^2)$ part of $\phi$, while the first post-Newtonian order is given by the square of the Newtonian terms, the $O(\epsilon^2)$ part of $\psi$, the $O(\epsilon^3)$ part of $A_i$, and the $O(\epsilon^4)$ part of $g_{00}$. The lowest non-trivial order in the equations of motion of massless test particles is given by the $O(\epsilon^2)$ parts of both $\phi$ and $\psi$, which are therefore sufficient to determine the post-Newtonian effects of gravity on rays of light.

In Section \ref{sec:pn} we will construct a cosmological model by gluing together different regions of space-time, each of which is modelled using the post-Newtonian expansion described above. This will follow the methodology used in Ref. \cite{Clifton}.

\section{The Top-Down Approach: \newline FLRW cosmologies}
\label{sec:flrw}

What we call the `top-down approach' to cosmology is, in fact, the usual way that cosmological models are constructed in the bulk of the literature. This approach starts by assuming that on sufficiently large scales the geometry of the Universe looks, in some sense, close to being spatially homogeneous and isotropic, so that it can be modelled using a Robertson-Walker geometry. It is then further assumed that this geometry should satisfy a given set of gravitational field equations, such as  Eqs. (\ref{fried})-(\ref{cons:perfect}). The former of these assumptions appears to be compatible with the majority of cosmological observations, but the latter is a purely mathematical question. Proving its validity requires proving that a local theory of gravity, such as $f(R)$ gravity, is applicable to a non-local averaged, or approximate, geometry. There are a number of extremely difficult technical questions that remain to be answered in order to rigorously justify such an approach \cite{av1,av2}. Nevertheless, modelling the Universe in this way appears to be able to account for the vast majority of cosmological observations with only two unknown matter fields. It is also an extremely simple way to model the Universe, and is used almost ubiquitously. In this section we will therefore briefly review the FLRW cosmological models of $f(R)$ theories of gravity, emphasizing in particular the aspects that are of most relevance for the bottom-up constructions we will study in Section \ref{sec:pn}.


\subsection{The Reconstruction Method}
\label{sec:recon}

This approach assumes that the expansion history of the universe is known exactly, and then inverts the field equations to deduce what classes of $f(R)$ theories give rise to this particular cosmological evolution \cite{reconstruction}-\cite{reconstruction2}. It is useful for demonstrating the freedom that exists in the Friedmann solutions of $f(R)$ theories.

The reconstruction method works as follows. If we presume we know $a=a(t)$, then we can take the usual expression for the Ricci curvature scalar, $R= 6 \dot{H} + 12 H^2,$ and invert it to find an expression for $t$ as a function of $R$:
\be
t = g_1(R) \, .
\ee
This function can then be used to write all expressions, that were previously functions of $t$, as functions of $R$. 
The Friedmann equation (\ref{fried}) and conservation equation (\ref{cons:perfect}) then give
\ba
\nonumber
&&3H [g_1(R)] \dot{R}[g_1(R)] f'' + f' \left( 3 H[g_1(R)]^2 - \frac{R}{2} \right) + \frac{f}{2} \\&=&\rho_0 a[g_1(R)]^{-3 (1+w)} \, ,
\ea
where $\rho_0$ is a constant, and where $w=p/\rho$ has been taken to be the constant equation of state of the perfect fluid. This is a second-order differential equation
that can be used to find the $f(R)$ that produce the stated $a(t)$. It works for almost any expansion history. Similar methods can be used  to reconstruct $f(R)$ from the functions $\dot{a}=g_2(a)$, $\dot{H}=g_3(H)$ and $\dot{q}=g_4(q)$ (where $q \equiv-\ddot{a} a/\dot{a}^2$ is the deceleration parameter) \cite{reconstruction2}. 

The freedom to produce such a wide array of different expansion histories has led to considerable interest in $f(R)$ gravity as a possible explanation of dark energy. In Section \ref{sec:pn} we will use bottom-up constructions to show that this freedom does not in fact exist, if the theory is required to simultaneously produce an observationally viable weak-field limit on small scales. All viable theories must expand like $\Lambda$CDM.

\subsection{Numerical Methods}

Numerical methods can also be used to investigate the Friedmann solutions of $f(R)$ theories. Of particular interest in this regard is the theory proposed by Hu $\&$ Sawicki \cite{Sawicki}. This theory has proven to be extremely popular over the past few years, and is generated from the function 
\be
f(R)\,=\, R-R_0\frac{b\left(\frac{R}{R_0}\right)^n}{1+d\left( \frac{R}{R_0}\right)^n}\,,
\label{HS}
\ee
where $R_0$, $b$, $d$ and $n$ are all positive constants. The constant $R_0$ has the dimensions of the Hubble parameter squared, and corresponds to the scale at which infrared modifications start to become important.  Because both $f'$ and $f''$ are positive, the theory has a positive effective gravitational constant and does not initially appear to posses any obvious instabilities. On the surface, this theory (and others like it \cite{Appleby,Starobinsky}) therefore appear to represent viable descriptions of a cosmic acceleration at late-times, driven by nothing more than the curvature of space-time.


\begin{figure}[htbp!]
\vspace{-1.5cm}
\includegraphics[width=0.7\textwidth]{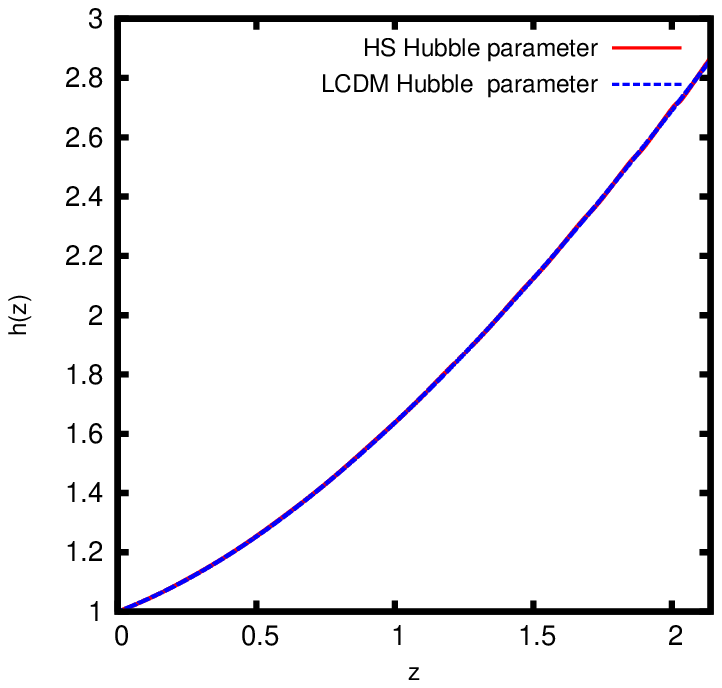}

\vspace{-1.5cm}
\includegraphics[width=0.7\textwidth]{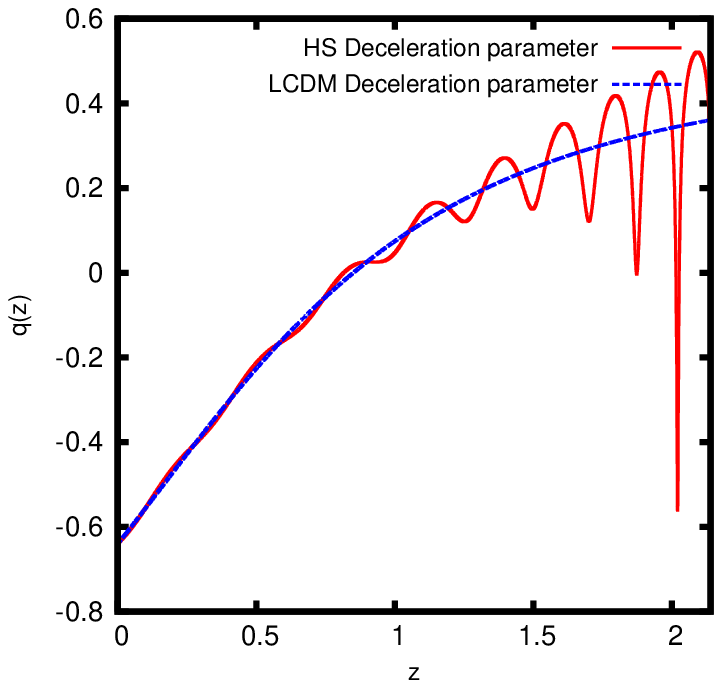}

 \vspace{-1.5cm}
\includegraphics[width=0.7\textwidth]{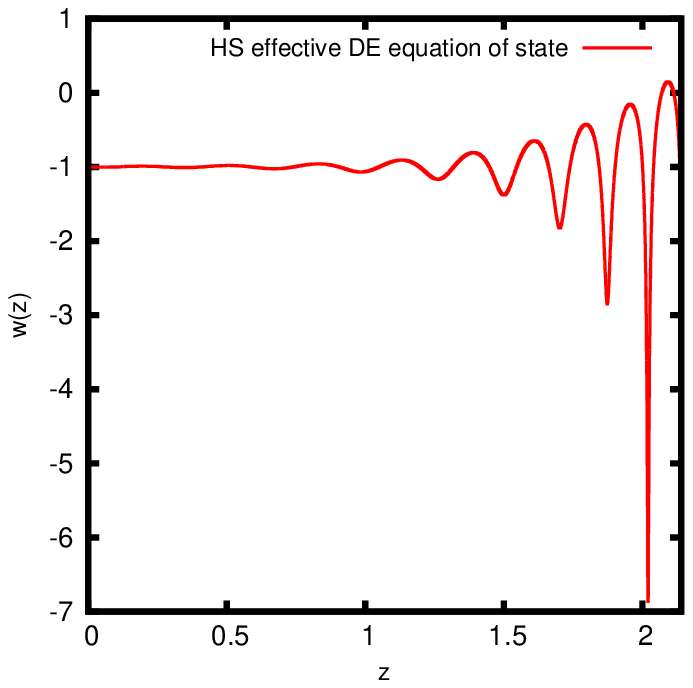}\\
\caption{\footnotesize{ 
The Hubble, deceleration and effective equation of state of dark energy $w_R$ defined by equation (\ref{DE}), for the Hu $\&$ Sawicki model with $n=3$. Here $b=1$, $d=0.1$ and we have taken $\Omega_{\Lambda}=0.76$. The condition $R_0b/d=6\Omega_{\Lambda}H_0^2$ gives $R_0=0.456H_0^2$.}}
 \label{singularity}
\end{figure}

An example of the numerical solutions to Eqs. (\ref{fried})-(\ref{cons:perfect}) for the Hu $\&$ Sawicki theory (\ref{HS}) with $n=3$ is presented in Fig. \ref{singularity}. In producing these plots we have assumed that $H(z)$ and $q(z)$ take on their $\Lambda$CDM values at the present day ($z=0$), and have chosen $R_0 b/d=6\Omega_{\Lambda}H_0^2$.
While the Hubble parameter as a function of redshift is very close to the $\Lambda$CDM value, it is clear that growing oscillations are present in the deceleration parameter, and that these oscillations translate into an effective equation of state for dark energy that varies wildly. In fact, it turns out that these oscillations lead to curvature singularities at finite times.
This instability was pointed out by Frolov \cite{Frolov}, and has since been studied numerically \cite{Tsujikawa:2007xu, Appleby:2008tv}.

These oscillations represent a severe drawback for the viability of such theories. Their suppression requires a careful fine-tuning of the parameters of the model, on top of what is required to set the value of the effective cosmological constant. On the other hand, they potentially provide a way of testing these $f(R)$ theories against supernovae and growth rate measurements
\cite{sulona2}. A further difficulty for the $n=1$ case, with the same initial conditions as above, is that the matter dominated epoch at high redshifts does not even exist \cite{sulona1}. 

In Section \ref{sec:pn} we will show that these oscillations do not occur at leading order in  the bottom-up constructions, as long as the slow-motion weak-field limit holds on small scales. This shows that, for perfectly homogeneous and isotropic FLRW models, the post-Newtonian limit must be destroyed on small scales when the oscillations start. However, it also suggests the intriguing possibility that the growth of structure in the real Universe could suppress the unstable oscillations. This is because non-linear structures keep the Ricci curvature scalar (which is a single-valued locally-defined quantity) in the high-curvature regime at all points in space. A corollary is then that the expansion histories of universes that obey such theories must be constrained to be even closer to the $\Lambda$CDM model than had previously been thought. We will return to these points in Section \ref{disc}.

\section{Bottom-Up Constructions: \newline A Patchwork Universe}
\label{sec:pn}

What we mean by `bottom-up constructions' are cosmological models in which the geometry of space-time in the vicinity of astrophysical objects is the thing that is solved for first, and given primary importance. This can usually be done up to the application of suitable boundary conditions, for the partial differential equations involved. Such boundary conditions are imposed by patching together a large number of sub-horizon sized regions of space-time using the Israel junction conditions. These same junction conditions simultaneously specify the motion of the boundaries between regions, and hence can be used to extrapolate the large-scale cosmological expansion of the space-time as a whole. In this approach to cosmological modelling, the field equations need only be applied locally (i.e., not to any non-locally averaged geometries), and the cosmological expansion can therefore be viewed as an emergent phenomenon, as it should be. The approach we will use here was originally proposed for GR in \cite{Clifton}, and was further developed in \cite{Clifton2}. It was adapted to the study of some $f(R)$ theories of gravity in \cite{ref1}. Here we will further adapt it to study the emergence of accelerating expansion in $f(R)$ gravity.

The situations we will primarily be concerned with are ones in which matter is clumped into highly non-linear structures on small scales. We will then take a sub-horizon sized region of space, which we will call a ``cell''. We have in mind that these cells should be polyhedral. We will treat the geometry of space-time inside each of these cells as being well approximated by the post-Newtonian expansion, as outlined in Section \ref{pn}. We will then consider many such cells, all the same shape, that are invariant under discrete rotations, and that are patched together about reflection symmetric boundaries. On very large scales the global geometry of space-time can then be thought of as statistically homogeneous and isotropic, while being very inhomogeneous and anisotropic on small scales. At no point in this process do we assume a global Robertson-Walker background.

While this approach is quite flexible, for definiteness we could consider each cell to contain a galaxy-sized amount of mass ($\sim 10^{12}$ solar masses), and to have a spatial extent that is similar to the inter-galactic separation in the actual Universe ($\sim$ 1 Mpc). If we then assume that the rate of expansion of our model Universe is of a similar order of magnitude to that of the real Universe ($\sim 70$ km  s$^{-1}$ Mpc$^{-1}$), then at the edge of each cell we will have a Hubble flow given by
\be
v \sim H  d \sim 2 \times 10^{-4} c \, .
\ee
The typical gravitational potential for structures of this size, separated by this distance is then
\be
\phi \sim \frac{G m}{c^2 d} \sim 5 \times 10^{-8} \, .
\ee
We therefore have
\be
\phi \sim \frac{v^2}{c^2} \sim \epsilon^2 \, ,
\ee
where $\epsilon \sim 10^{-4}$. This is sufficient to be able to apply a weak-field and slow-motion post-Newtonian expansion of the field equations. A similar process could be considered for clusters of galaxies, which as long as the average density is unchanged, would result in an expansion parameter of order $\epsilon \sim 10^{-3}$.

Crucially, when considering general $f(R)$ theories of gravity, we will not be able to make any assumptions about the smallness (or not), of $f$ and $f'$. We will also have to refrain from expanding these quantities around $R=0$, even though our perturbative expansion within each cell will be around Minkowski space. This is done so as to include as many $f(R)$ theories as possible, some of which contain parameters that are small, and some of which are only expected to admit convergent power series expansions around finite values of $R$. We will, however, have to assume that $f$ and $f'$ can be expanded in our order-of-smallness parameter $\epsilon$ (if this is not true then the theories cannot be treated perturbatively at all). In this case we have
\be
f = f^{(0)} + f^{(2)} + O(\epsilon^4)
\ee
and
\be
f' = f'^{(0)} + f'^{(2)} + O(\epsilon^4) \, ,
\ee
where superscripts in brackets denote the order of smallness of a term in the $\epsilon$ expansion. As usual in post-Newtonian expansions, we have expanded these scalars in even powers of the expansion parameter $\epsilon$. The convergence of these expansions can be verified {\it a posteriori}.

The junction conditions we will use when joining cells together will be those found by Deruelle, Sasaki and Sendouda \cite{der} in the absence of surface layers, and when $f' \neq 0$ and $f''\neq0$:
\ba
\left[\gamma_{ab} \right]^+_- &=& 0
\\ \left[ K_{ab} \right]^+_- &=& 0
\\ \left[ R \right]^+_- &=& 0
\\ \left[ \partial_y f' \right]^+_- &=& 0 \; ,
\ea
where $\gamma_{ab}$ and $K_{ab}$ are the first and second fundamental forms on the boundary, where $a,b ...$ are indices on the $1+2$-dimensional boundaries, and where $\partial_y$ is a normal derivative on the boundary.

For the lattice construction described above these junction conditions provide Neumann boundary conditions for the field equations within the interior of each cell. This can be contrasted with Swiss cheese constructions, in which the boundary conditions imposed on these equations would be of Cauchy-type, and the imposition of ``cosmological boundary conditions'', which are often taken to be of Dirichlet-type (with the idea that weak-field solutions slowly turn into cosmological solutions at very large distances).


The geometry that will be used will be given by Eq. (\ref{pnmetric}). Initially, it may seem that the zero-order equations here are completely trivial (i.e., that they are immediately solved as long as Minkowski space is a solution of the theory). However, we should bear in mind that many of the theories of interest to us at present are those that have been constructed to have interesting behaviour in $f'$ when $R$ is small. We therefore cannot immediately assume anything about the smallness of the lowest order non-zero parts of $f$ and $f'$, but instead must determine their magnitude through the field equations (\ref{field}). 

From the zero-order part of the field equations it can be seen from $T^{(0)}_{00}+T^{(0)}_{xx}=0$, $T^{(0)}_{00}+T^{(0)}_{yy}=0$ and $T^{(0)}_{00}+T^{(0)}_{zz}=0$ that we must have
\be
\label{fpxx}
f'^{(0)}_{,xx} =  f'^{(0)}_{,yy} = f'^{(0)}_{,zz} = 0.
\ee
The equations $T^{(0)}_{xy}=T^{(0)}_{xz}=T^{(0)}_{yz}=0$ also give
\be
\label{fpxx2}
f'^{(0)}_{,xy} =  f'^{(0)}_{,xz} = f'^{(0)}_{,yz} = 0.
\ee
Substituting this into the $T^{(0)}_{00}=0$ equation then gives
\be
f^{(0)}=0\;,
\ee
so that the leading order term in $f$ must be $O(\epsilon^2)$ or smaller. The leading order term in $f'$, on the other hand, must be given by the integral of Eqs. (\ref{fpxx}) and (\ref{fpxx2}):
\be
f'^{(0)} = \alpha (t) + \beta_1(t) x +\beta_2(t) y + \beta_3 (t) z \, ,
\label{fp1}
\ee
where $\beta_1$, $\beta_2$, $\beta_3$, and $\alpha$ are arbitrary functions of $t$. However, it can immediately be seen from the junction condition $\left[ \partial_y f' \right]^+_- = 0$, and the imposition of reflection symmetric boundaries, that we must have $\partial_y f'=0$ at every point on the edge of every cell. Using Eq. (\ref{fp1}), this means that we must have $\beta_1=\beta_2=\beta_3=0$. 

The leading order behaviour of $f'$ must therefore be spatially homogeneous, so that
\be
f' = \alpha (t) + O(\epsilon^2) \, .
\ee

This result suggests that we can consider two separate classes of theory:
\begin{itemize}
\item{\bf Class I:} Theories that can be written as $f(R)=\alpha R + F(R)$,  such that $f'=\alpha + F'$, where $\alpha$ is a constant (not a function of $R$).
\item{\bf Class II:} Theories with $R=R_0 +O(\epsilon^4)$, where $R_0=R_0(t) \sim \epsilon^2$.
\end{itemize}

In Class I we have all theories in which $f'^{(0)}$ is not a function of $R$, but is instead constructed from constant parameters that appear in the gravitational Lagrangian. In this case we can integrate to find $f(R)=\alpha R+F(R)$, such that $F \sim F'\sim \epsilon^2$. The Class I solutions contain most of the viable $f(R)$ theories that  have recently been considered in the literature (e.g. Hu \& Sawicki \cite{Sawicki}, Appleby \& Battye \cite{Appleby}, and Starobinsky \cite{Starobinsky}).

In Class II we have all theories in which $f'^{(0)}$ is a non-trivial function of the Ricci curvature scalar, $R$. If this is true, together with the fact that $f'$ is a function of $t$ only, then the lowest order part of $R$ must also be a function of $t$ only. In this case we can write $R = R_0 + O(\epsilon^4)$, where $R_0 =R_0(t)\sim \epsilon^2$. Any space dependent contributions to $R^{(2)}$ would not be compatible with the supposition $\alpha (t)=\alpha (R)$.

Let us now consider each of these two classes individually.

\subsection{Class I Theories}

In this case the lowest-order parts of $F$ and $F'$ are both $ O( \epsilon^2)$. The equations $T^{(2)}_{00}=8 \pi \rho$ and $T^{(2)}=-8 \pi \rho$ then give, to lowest order,
\ba
\label{dpsia1}
\alpha \nabla^2 \psi + \frac{1}{2} \nabla^2 F' &=& - 4 \pi \rho + \frac{1}{4} F\\
\alpha \nabla^2 \phi - \frac{1}{2} \nabla^2 F' &=& -4 \pi \rho -\frac{1}{2} F,
\label{dphia1}
\ea
where $\nabla^2$ denotes a spatial Laplacian operator. Likewise, the $T^{(2)}_{xy}=T^{(2)}_{xz}=T^{(2)}_{yz} = 0$ equations give
\ba
F'_{,xy} + \alpha \psi_{,xy} - \alpha \phi_{,xy} &=& 0\\
F'_{,xz} + \alpha \psi_{,xz} - \alpha \phi_{,xz} &=& 0\\
F'_{,yz} + \alpha \psi_{,yz} - \alpha \phi_{,yz} &=& 0.
\ea
These latter equations require
\be
\label{fra1}
F' = \alpha (\phi-\psi) + \frac{3}{4} \left( c_1(t,x) + c_2 (t,y) + c_3(t,z) \right),
\ee
where $c_1$, $c_2$ and $c_3$ are (so far) arbitrary functions of $t$, with $x$, $y$ and $z$, respectively. Combining Eqs. (\ref{dpsia1}), (\ref{dphia1}) and (\ref{fra1}) gives
\be
\label{fa1}
F = \nabla^2 (c_1 +c_2 +c_3),
\ee
which in turn allows us to note that, because $F_{,x}=F' R_{,x} \sim \epsilon^4$, we also have at lowest order that
\be
c_{1,xxx} = 0 , \qquad c_{2,yyy}=0 \qquad {\rm and} \qquad c_{3,zzz}=0.
\ee
This tells us that $c_1$, $c_2$ and $c_3$ all contain terms up to quadratic order in their respective spatial arguments only. Equation (\ref{fa1}) then becomes
\be
\label{fconst}
F= F_0(t),
\ee
and allows Eqs. (\ref{dpsia1}) and (\ref{dphia1}) to be written as
\ba
\label{psia3}
\nabla^2 \left( \alpha \psi + \frac{1}{2} F' \right) &=& -4 \pi \rho +\frac{1}{4} F_0\\
\nabla^2 \left( \alpha \phi - \frac{1}{2} F'  \right) &=& -4 \pi \rho -\frac{1}{2} F_0.
\label{phia3}
\ea
At this point it is convenient to consider the $T^{(3)}_{0i} = -8\pi \rho v_i$ equations:
\be
\frac{1}{4} \alpha \nabla^2 A_i - \frac{1}{4} \alpha A_{j,ji}+
\alpha \dot{\psi}_{,i} +\frac{1}{2} \dot{F}^{\prime}_{,i} = 4\pi \rho v_i.
\ee
Substituting this into the time-derivative of Eq. (\ref{psia3}) then gives
\be
4 \pi (\rho v_i)_{,i} = -4 \pi \dot{\rho} +\frac{1}{4} \dot{F}_0,
\ee
which on using the Eulerian continuity equation of hydrodynamics for a perfect, non-viscous fluid gives $$F_0= {\rm constant} \, .$$

Equations (\ref{psia3}) and (\ref{phia3}) can now be seen to have the solutions
\ba
\label{psia2}
\psi &=& \frac{1}{\alpha} \left( U - \frac{1}{2} F' \right)\\
\phi &=& \frac{1}{\alpha} \left( \hat{U} + \frac{1}{2} F' \right),
\label{phia2}
\ea
where $U$ and $\hat{U}$ denote the usual Newtonian potentials, including cosmological constant, and are given implicitly by the following expressions:
\ba
\nabla^2 U &=& -4 \pi \rho + \frac{1}{4} F_0\\
\nabla^2 \hat{U} &=& -4 \pi \rho - \frac{1}{2} F_0.
\ea

Now, the large-scale expansion of the cosmological model obtained from patching together our weak-field regions is given by \cite{Clifton}. The outward acceleration of each point on the boundary of every cell is
\be
\ddot{x} = {\partial_y \phi} +O(\epsilon^4) \, .
\ee
Or, defining the average outward expansion by
\be
\ddot{a} \equiv \frac{\int \ddot{x} dA}{\int dA} \, ,
\ee
where the integrals here are over the surface of a cell face, we can write
\be
\label{rayup}
\frac{\ddot{a}}{a} = - \frac{4 \pi G_{\rm eff}}{3} \frac{M}{\gamma_2 a^3} + \frac{\Lambda_{R}}{3} \, ,
\ee
where $G_{\rm eff} \equiv 3 \gamma_2/n \alpha \gamma_1$, where $M \equiv \int \rho dV$ is the integral of the energy density over the volume of the cell, where $n$ is the number of faces to a cell, and where
\be
\Lambda_R \equiv - \frac{ G_{\rm eff} F_0}{2}
\ee
is the contribution to the cosmological constant from the $F(R)$ part of the action. We have also defined the area of a cell face and the volume of the cell to be $A \equiv \gamma_1 a^2$ and $V \equiv \gamma_2 a^3$, where $\gamma_1$ and $\gamma_2$ are constants. Of course, if the cell maintains its shape during the evolution, as expected, then we have $\ddot{a}=\ddot{x}$. Notice that both $G_{\rm eff}$ and $\Lambda_R$ are constants, to the required order.

The parts of Eq. (\ref{rayup}) that look like dust and a cosmological constant come from the normal derivative of $\hat{U}$ on the boundary of the cell. The terms involving the normal derivative of $F'$, however, do not contribute to the large-scale evolution at lowest order at all, because the junction condition $\left[ \partial_y f' \right]^+_- = 0$ evaluated at $O(\epsilon^2)$ gives $\partial_y F'=0$. The bottom-up cosmological solutions of all theories in this class therefore behave like GR plus $\Lambda$, with corrections appearing at $O(\epsilon^4$) only.

The only new behaviour possible in this sub-class arises in the interior of each cell, and is contributed by the possible presence of an $O(\epsilon^2)$ term in $F'$. Any such term must obey the trace of the field equations:
\be
\label{scalara}
\nabla^2 F' = \frac{1}{3} R + \frac{2}{3} F_0 - \frac{8 \pi}{3} \rho.
\ee
The solutions to this equation may or may not obey a screening mechanism, but either way do not affect the leading-order behaviour of the large-scale expansion.

As $F_0$ behaves as the effective cosmological constant, and because $F_0$ must be constructed from constants that appear in the gravitational action only, this class of theories does not appear to solve any of the problems associated with the cosmological constant (i.e., why should $F_0$ cancel with the vacuum energy that appears on the right-hand side of field equations, and why should the value of $F_0$ be of the same order of magnitude as the energy density in matter at the present time?). Beyond this, one may also notice that there are no signs of the oscillations that appear in the FLRW solutions to these theories.

\subsection{Class II Theories}

In this class of theories we are unable to directly integrate $f'$ in order to find the lowest-order part of $f(R)$. We therefore resort to writing
\be
f = f^{(2)} + O(\epsilon^4)
\ee
and
\be
f' = \alpha + f'^{(2)} + O(\epsilon^4), 
\ee
where $\alpha=\alpha(t)\sim 1$. In this case, the equations $T^{(2)}_{00}= 8 \pi \rho$ and $2 T^{(2)}_{00} + T^{(2)} = 8\pi \rho$ give
\be
\label{phib2}
\frac{1}{2} f^{(2)} - \nabla^2 f'^{(2)} - \alpha \nabla^2 \phi = 8\pi \rho
\ee
and
\be
-f^{(2)} - 3 \ddot{\alpha} + \nabla^2 f'^{(2)} - 4 \alpha \nabla^2 \psi = 8 \pi \rho \, , 
\label{psib2}
\ee
and the $T^{(2)}_{ij}=0$ equations can be integrated to find
\be
f'^{(2)} = \alpha \left( \phi - \psi \right) + \frac{3}{4} (c_1(t,x) +c_2 (t,y) + c_3(t,z)),
\ee
where $c_1$, $c_2$ and $c_3$ are arbitrary functions of their arguments. These equations can then be manipulated to obtain
\ba
\label{psib1}
\alpha \nabla^2 \psi &=& - \frac{8 \pi}{3} \rho - \frac{1}{6} f^{(2)}- \frac{2}{3} \ddot{\alpha} \\&&\qquad+\frac{1}{12} \nabla^2 (c_1+c_2+c_3)  \nonumber
\ea
and
\ba
 \label{phib1} \alpha \nabla^2 \phi &=& - \frac{16 \pi}{3} \rho + \frac{1}{6} f^{(2)}- \frac{1}{3} \ddot{\alpha}\\&&\qquad- \frac{1}{3} \nabla^2 (c_1+c_2+c_3) \nonumber
.
\label{frb1}
\ea
From the definition of $R$ we also have
\ba
\nonumber
\alpha R &=& 2 \alpha \left( \nabla^2 \phi - 2 \nabla^2 \psi \right) \\ &=& f^{(2)} + 2 \ddot{\alpha} - \nabla^2  (c_1+c_2+c_3) \, .
\label{rb1}
\ea
Now, because we have in this class of theories that $R^{(2)}=R^{(2)}(t)$, we must also have that $f^{(2)}=f^{(2)}(t)$ (as this is the leading order term in $f$, which is a function of $R$ only). Equation (\ref{rb1}) then shows that $c_1$, $c_2$ and $c_3$ must be at most quadratic in $x$, $y$ and $z$, respectively, so that $\nabla^2  (c_1+c_2+c_3)$ is a function of $t$ only.

Using this information, Eqs. (\ref{psib1}) and (\ref{phib1}) can be written
\be
\alpha \nabla^2 \psi = - \frac{8 \pi}{3} \rho + {\rm homogeneous~terms}
\ee
and
\be
\alpha \nabla^2 \phi = - \frac{16 \pi}{3} \rho + {\rm homogeneous~terms} \; ,
\ee
which have the solutions
\be
\label{f1}
\psi = \frac{2 \bar{U} }{3 \alpha} + {\rm cosmological~terms}
\ee
and
\be
\phi = \frac{4 \bar{U}}{3 \alpha}  + {\rm cosmological~terms} \; , \label{f2}
\ee
where $\bar{U}$ is the Newtonian potential that satisfies $$\nabla^2 \bar{U} = - 4 \pi \rho \, ,$$ and where ``cosmological terms'' refers to potentials that are independent of $\rho$, and proportional to $r^2$. It can immediately be seen that all theories in this class have the post-Newtonian parameter 
\be
\gamma_{\rm PPN} \equiv \frac{\psi}{\phi} = \frac{1}{2} \; ,
\ee
where cosmological terms have been neglected. The Chameleon Mechanism cannot apply to this class of theories, as it requires $R$ to be inhomogeneous for it to be effective. This can be seen from the trace of the field equations, which can be written as
\ba
\nabla^2 f'^{(2)} &=& -\frac{8 \pi}{3} \rho + \frac{1}{3} \alpha R +\frac{2}{3} \nabla^2 (c_1+c_2+c_3) \\&=& -\frac{8 \pi}{3} \rho + {\rm homogeneous~terms}  \nonumber \, .
\ea
The solution for $f'^{(2)}$ is then simply
\be
f'^{(2)} = \frac{2 \bar{U}}{3} + {\rm cosmological~terms} \, ,
\ee
which does not permit the type of non-trivial distributions that are required to implement the Chameleon mechanism, and suppress the propagation of the scalar degree of freedom.

The fact that $\gamma_{\rm PPN}=1/2$ is also evident directly from the definition of $R$, which gives 
\ba
R &=& 2 \nabla^2 \phi - 4 \nabla^2 \psi \\ &=& - \frac{32 \pi}{3 \alpha} (1-2 \gamma_{\rm PPN}) \rho + {\rm homogeneous~terms} \, .
\nonumber
\ea
For this equation to be simultaneously consistent with $R=R(t)$, and an inhomogeneous $\rho=\rho(t,x,y,z)$, we clearly require $\gamma_{\rm PPN} = 1/2$.

We therefore find that theories in this class, while potentially admitting new and interesting Friedmann solutions, cannot be considered observationally viable, as they conflict with experimental probes of post-Newtonian gravity, which currently require \cite{shapiro}
\be
\gamma_{\rm PPN} -1 = (-1.7 \pm 4.5) \times 10^{-4} \, .
\ee
This is inconsistent with $\gamma_{\rm PPN} = 1/2$ at more than $1000 \sigma$.

\section{Discussion}
\label{disc}

We have studied the ability of $f(R)$ theories of gravity to account for the present day accelerated expansion of the Universe using the top-down and bottom-up approaches to cosmological modelling. The top-down approach admits a wide variety of possible evolutions. They also tend to admit oscillations about background expansions that are close to that of the $\Lambda$CDM model. These oscillations are a potential way to distinguish between $f(R)$ cosmology and the $\Lambda$CDM model, but they also appear to signify an instability within the FLRW models of $f(R)$ theories.

Within the bottom-up constructions we identified two possible classes of theory, which we named Class I and Class II. The Class I theories are potentially observationally viable, and include most of the well-studied theories that have recently appeared in the literature \cite{Sawicki}-\cite{Starobinsky}. The expansion history that emerges within this class of theories is indistinguishable from the $\Lambda$CDM model, and is not plagued by the potentially problematic oscillations that are known to occur in the Friedmann solutions of these theories. This appears to be a type of strong back-reaction, and signifies that either the growth of structure suppresses the oscillations that occur in the homogeneous and isotropic solutions, or that the occurrence of oscillations destroys the $\epsilon$ expansion that we have used to describe the gravitational field on small scales. This could mean that either the fluctuations become large or rapidly changing, so that the $\epsilon$ expansion is no longer convergent, or that the Minkowski space background becomes unsuitable. In either case the weak-field slow-motion treatment of gravitational fields must be said to be no longer applicable. If this is the case then it is extremely difficult to see how the theory could be considered to be observationally viable.

The Class II theories have the potential to describe new and interesting large-scale behaviour, due to the new cosmological terms in Eqs. (\ref{f1}) and (\ref{f2}). However, it can be seen that they must all have a weak-field limit that is not compatible with observations of post-Newtonian gravity, as they have $\gamma_{\rm PPN}=1/2$. The Chameleon mechanism cannot apply to this class of theories, as the lowest-order part of $R$ must be a function of time only. So, while theories in this class are potentially capable of displaying a variety of cosmological expansion histories, we are led to conclude that they are not observationally viable on all scales.

Our key result is therefore the following: All $f(R)$ theories that are potentially compatible with observations have expansion histories in the late Universe that are indistinguishable from GR with $\Lambda$. Moreover, the value of the effective cosmological constant in these models must be a combination of the bare cosmological constant, the vacuum energy density ($\rho_{\rm vac}$), and a possible combination of other constant parameters that appear in the gravitational action of the $f(R)$ theory in question:
\be
\Lambda_{\rm effective} = \Lambda_{\rm bare} - \frac{4 \pi}{\alpha} \rho_{\rm vac} + \Lambda_R \, .
\ee
The fine-tuning that is required to get $\Lambda_{\rm effective}$ at the level required in order to explain observations is therefore not at all alleviated by generalising the gravitational Lagrangian from the Einstein-Hilbert one to the $f(R)$ family. To say that theories in this class cause accelerating expansion would also appear to be misleading: The only parts of the gravitational Lagrangian that are contributing to the acceleration at late times are those parts that, to leading order, are constants.

A natural question that arises is whether the results found above also apply to scalar-tensor theories of gravity, as the $f(R)$ theories are known to be equivalent to a class of these theories after a Legendre transformation. To be more explicit, if we make the definition
\be
\varphi \equiv f' \, ,
\ee
and if we assume $\varphi = \varphi (R)$ is invertible, so that we can make the further definition
\be
V(\varphi) \equiv \frac{1}{2} \left[ R(\varphi) \varphi - f(R(\varphi)) \right] \, ,
\ee
then we are led to the following gravitational Lagrangian:
\be
\mathcal{L}= \frac{1}{16 \pi} \left[ \varphi R - 2 V(\varphi) \right] \, .
\ee
This is the sub-class of scalar-tensor theories that have vanishing coupling constant, $\omega=0$. In this case the field equations can be written
\be
\label{STfe}
\varphi G_{\mu \nu} + (\nabla^{\sigma} \nabla_{\sigma} \varphi + V) g_{\mu \nu} -  \nabla_{\mu} \nabla_{\nu} \varphi = 8 \pi T_{\mu \nu} \, ,
\ee
and the scalar field propagation equation becomes
\be
\label{STs}
\nabla^{\sigma} \nabla_{\sigma} \varphi = \frac{8 \pi}{3} T - \frac{4}{3} V + \frac{2}{3} \varphi \frac{dV}{d\varphi} \, .
\ee
Taking the trace of Eq. (\ref{STfe}), and using Eq. (\ref{STs}), then gives
\be
R= 2 \frac{dV}{d\varphi} \, .
\ee
This is an algebraic relationship between the Ricci curvature scalar of the space-time, and the scalar field in the gravitational theory. It shows that in scalar-tensor theories with $\omega=0$ and $V(\varphi )\neq$~constant that if we know information about the curvature of the space-time (by assuming it can be described by a weak-field expansion on small scales, for example), then we also know something about the scalar degree of freedom in the theory. This was a key point used in finding the results above for $f(R)$ theories of gravity, and so we also expect our results to apply to the $\omega=0$ sub-class of scalar-tensor theories. This does not, however, mean that they should be expected to apply to scalar-tensor theories in general, as there is in general no algebraic relation between $R$ and $\varphi$.

\section*{Acknowledgements}

TC is supported by the STFC. PKSD is supported by the NRF (South Africa).


\end{document}